\documentstyle[prb,aps]{revtex}

\begin{document}
\title{Inelastic co-tunneling through an excited state of a quantum dot}
\author{M. R. Wegewijs, Yu. V. Nazarov}
\address{Department of Applied Physics, Faculty of Applied Science,\\
Delft University of Technology, Lorentzweg 1, 2628 CJ Delft,The Netherlands}
\maketitle

\begin{abstract}
We consider the transport spectroscopy of a quantum dot with an even number
of electrons at finite bias voltage within the Coulomb blockade diamond. We
calculate the tunneling current due to the elastic and inelastic
co-tunneling processes associated with the spin-singlet ground state and the
spin-triplet first excited state. We find a step in the differential
conductance at a bias voltage equal to the excitation energy with a peak at
the step edge. This may explain the recently observed sharp features in
finite bias spectroscopy of semi-conductor quantum dots and carbon
nanotubes. Two limiting cases are considered: (i) for a low excitation
energy, the excited state can decay only by inelastic co-tunneling due to
Coulomb blockade (ii) for a higher excitation energy the excited state can
decay by sequential tunneling. We consider two spin-degenerate orbitals that
are active in the transport. The nonequilibrium state of the dot is
described using master equations taking spin degrees of freedom into
account. The transition rates are calculated up to second order in
perturbation theory. To calculate the current we derive a closed set of
master equations for the spin-averaged occupations of the transport states.
\end{abstract}

\section{Introduction}

Transport spectroscopy provides information on the excited states of quantum
dots.~\cite{bib:review} Outside the Coulomb blockaded regions
(``diamonds''), peaks in the differential conductance can be observed, which
linearly shift with the gate voltage.~\cite{bib:excpeaks} Such a peak
indicates that an excited state has entered the transport voltage window.
Recently, De Franceschi et al. observed new features in the spectroscopy
data of a vertical semiconductor quantum dot coupled relatively strong to
the leads.~\cite{bib:SCdot1} {\em Inside} several Coulomb diamonds with an
even number of electrons they observed a step in the differential
conductance with a peak at the step edge for a quantum dot with six
electrons. In contrast to the peaks outside the diamond, this feature
depends very weakly on the gate-voltage. This was interpreted as being due
to {\em inelastic co-tunneling} through an excited state of the dot. It was
pointed out that the sharpness of the step is determined either the
temperature or by the broadening of the excited level, both of which are
small energy scales compared to the widths of the first-order tunneling
peaks. This provides an new spectroscopic tool for studying excited levels
in quantum dots. Similar features were observed in the transport
spectroscopy of carbon nanotubes obtained by Nyg\aa rd et. al..~\cite{bib:NT}
Here also, the gate voltage independent features at finite bias voltage were
clearly seen in consecutive Coulomb diamonds with an even number of
electrons. The observed peaks at the step edges were attributed to
Kondo-physics. In both experiments a Kondo effect was observed at zero bias
voltage in a magnetic field. The occurence of Kondo peaks in the
differential conductance, although of different origin, indicates a
relatively strong coupling to the leads.~\cite
{bib:Kondo,bib:Kondo2,bib:n=evendots}

Motivated by these experiments, we have calculated the co-tunneling current
through an excited level of a quantum dot using a master equation approach.
We investigate the possibility that the observed features arise from changes
in the nonequilibrium occupations of the states in the dot. Such changes are
caused by voltage-dependence of co-tunneling processes. We find that a peak
structure in the differential conductance can develop at the onset of
inelastic co-tunneling. We consider the possible regimes defined by the
energy of the excited level relative to the addition energy. Two possible
cases are distinguished:~\cite{bib:SCdot1} (i) for low excitation energy,
the excited state can decay only by inelastic co-tunneling due to Coulomb
blockade; (ii) for higher excitation energy the excited state can decay by
sequential tunneling. We take the spin degrees of freedom into account and
derive master equations for the nonequilibrium occupations of the relevant
states, which are averaged over the spin projections. The case (i) has
recently been considered by M. Eto~\cite{bib:Eto} without taking the spin
degrees of freedom into account.

The plan of the paper is as follows. In Sec.~\ref{sec:transp} we introduce
a model for the transport spectroscopy of a quantum dot. The low bias
elastic co-tunneling current is discussed in Sec.~\ref{sec:el} and in Sec.~%
\ref{sec:inel} we consider the different regions of the Coulomb diamond at
higher bias where inelastic co-tunneling dominates the transport. Our
conclusions are presented in Sec.~\ref{sec:concl1}.

\section{Transport spectroscopy}

\label{sec:transp}We consider a quantum dot connected to two leads $L$ and $%
R $ and one gate electrode $G$. The dot is modeled by the Hamiltonian $H=H_{%
\text{sp}}+H_{\text{el}}+H_{\text{ex}}$. The discrete single particle states
are labeled by $d=1,2,\ldots $ and are spin degenerate, $s=\uparrow
,\downarrow $: 
\[
H_{\text{sp}}=\sum_{ds}\varepsilon _{d}n_{ds}, 
\]
where $n_{ds}=a_{ds}^{\dagger }a_{ds}$ is the number operator of level $ds$.
We describe the electron-electron interactions in the classical electron
liquid picture (Coulomb blockade model)~\cite{bib:review}. The electrostatic
energy depends on the total number of electrons $N=\sum_{ds}n_{ds}$ in the
dot and on the voltages on the electrodes, which couple to the island
through the capacitances $C_{L},C_{R},C_{G}$ (total capacitance $%
C=\sum_{k=L,R,G}C_{k}$): 
\[
H_{\text{el}}=\frac{1}{2}N\left( N-1\right) U+\left( \frac{1}{2}%
U-\sum_{k=L,R,G}\frac{C_{k}}{C}V_{k}\right) N. 
\]
Here $U=1/C$ is the charging energy (we use units $e=k=\hbar =1$). We also
include an exchange term that lowers the energy of states with parallel
spins: 
\[
H_{\text{ex}}=-J\sum_{dd^{\prime }s}n_{ds}n_{d^{\prime }s} 
\]
The leads are modeled as noninteracting quasi-particle reservoirs with spin
degenerate states. For the left lead with states $l$ we write $%
H_{L}=\sum_{ls}\varepsilon _{l}n_{ls}$ where $n_{ls}=a_{ls}^{\dagger }a_{ls}$
and $H_{R}$ is defined similarly for the right lead (states labeled by $r$).
The leads are assumed to be in equilibrium at chemical potential $\mu
_{L}=\mu _{R}+V$ and $\mu _{R}$, respectively. They are coupled to the dot
by tunnel junctions. The tunneling Hamiltonian of the left junction is 
\[
H_{TL}=\sum_{ds}\left( H_{TLd}^{\dagger }+H_{TLd}\right) ,\quad
H_{TLd}=\sum_{ls}t_{ld}a_{ls}^{\dagger }a_{ds}, 
\]
and $H_{TR}$ is defined similarly for the right junction ($L,l\rightarrow
R,r $). The temperature $T$ is assumed to be small: $T\ll \Gamma ,V,\delta ,$
where $\Gamma $ is the typical broadening of an energy level in the dot due
to the coupling to the leads and $\delta $ is the relevant level spacing
introduced below.

We consider here the case where the number of electrons on the dot $N$ is 
{\em even} at zero bias voltage. Let us first consider the conditions for 
{\em sequential tunneling} transport through the dot. For transitions
between the $N$ and $N\pm 1$ electron ground states only two orbitals need
to be considered. Let $|0\rangle $ denote the $N-2$ electron ground state
and $d$ and $d^{\prime }$ label the $N/2^{\text{th}}$ and $N/2+1^{\text{st}}$
orbital with level spacing $\delta =\varepsilon _{N/2+1}-\varepsilon _{N/2}$%
. The singlet ground state then reads $|N_{0,0}\rangle =a_{d\uparrow
}^{\dagger }a_{d\downarrow }^{\dagger }|0\rangle $. Here the subscripts in $%
|N_{S,S_{z}}\rangle $ indicate the spin $S$ on the dot and its projection $%
S_{z}$ in some arbitrary $z$-direction. The doublet ground states with $N\pm
1$ electrons are denoted by%
\begin{mathletters}%
%
\label{eq:gsN-1N+1} 
\begin{eqnarray}
|\left( N-1\right) _{1/2,+1/2}\rangle &=&a_{d\uparrow }^{\dagger }|0\rangle
,\quad |\left( N+1\right) _{1/2,+1/2}\rangle =a_{d^{\prime }\uparrow
}^{\dagger }a_{d\uparrow }^{\dagger }a_{d\downarrow }^{\dagger }|0\rangle ,
\\
|\left( N-1\right) _{1/2,-1/2}\rangle &=&a_{d\downarrow }^{\dagger
}|0\rangle ,\quad |\left( N+1\right) _{1/2,-1/2}\rangle =a_{d^{\prime
}\downarrow }^{\dagger }a_{d\uparrow }^{\dagger }a_{d\downarrow }^{\dagger
}|0\rangle ,
\end{eqnarray}
\end{mathletters}%
%
The differences between the ground state energies $E_{N}$ are%
\begin{mathletters}%
%
\begin{eqnarray*}
E_{NN-1} &\equiv &E_{N}-E_{N-1}=\varepsilon _{N/2}+\left( N-\frac{1}{2}%
\right) U-\sum_{k=LRG}\frac{C_{k}}{C}V_{k}, \\
E_{N+1N} &\equiv &E_{N+1}-E_{N}=\varepsilon _{N/2+1}+\left( N+\frac{1}{2}%
\right) U-\sum_{k=LRG}\frac{C_{k}}{C}V_{k}.
\end{eqnarray*}
\end{mathletters}%
%
In order to have sequential tunneling transport it must be possible to
inject an electron from the left lead and emit one to the right lead in a
subsequent event:%
\begin{mathletters}%
%
\label{eq:window} 
\begin{eqnarray}
\mu _{L} &\geq &E_{NN-1}\geq \mu _{R}  \label{eq:window_NN-1} \\
\mu _{L} &\geq &E_{N+1N}\geq \mu _{R}  \label{eq:window_N+1N}
\end{eqnarray}
\end{mathletters}%
%
These transport conditions translate into a stability region in the plane of
bias and gate voltages, which is shown in Fig.~\ref{fig:spectr}. Let us
express the addition energies as a function of the position ($V,V_{G}$) in
this diagram. In the limit of zero bias voltage (i.e. $V=0$ in Fig.~\ref
{fig:spectr}), sequential tunneling transport is possible only at two
isolated points: $V_{G}=V_{G}^{NN-1}$ [Eq. (\ref{eq:window_NN-1}) is
satisfied] and $V_{G}=V_{G}^{N+1N}$ [Eq. (\ref{eq:window_N+1N}) is
satisfied]. At these degeneracy points the number of electrons on the dot
can change by one. At the intermediate gate voltages $%
V_{G}^{NN-1}<V_{G}<V_{G}^{N+1N}$ the number of electrons on the dot is
stable and equal to $N$. Sweeping the gate voltage one thus observes the
Coulomb blockade oscillations with peak spacing 
\[
V_{G}^{N+1N}-V_{G}^{NN-1}=\left( C/C_{G}\right) \left( \delta +U\right) . 
\]
This involves the level spacing $\delta $ because the number of electrons on
the dot is even. For a finite, fixed bias voltage each of these degeneracy
points develops into a gate-voltage transport window, cf. Fig.~\ref
{fig:spectr}. Equivalently, the transport conditions (\ref{eq:window_NN-1})
and (\ref{eq:window_N+1N}) require that the bias voltage exceeds a
gate-voltage dependent threshold to allow the number of electrons on the dot
to fluctuate. The threshold lines are (depicted in Fig.~\ref{fig:spectr})%
\begin{mathletters}%
%
\label{eq:Vth} 
\begin{eqnarray}
V_{L}^{NN-1} &\equiv &-%
{\displaystyle{C_{G} \over C-C_{L}}}%
\left( V_{G}-V_{G}^{NN-1}\right) ,  \label{eq:VthLNN-1} \\
V_{R}^{NN-1} &\equiv &+%
{\displaystyle{C_{G} \over C_{L}}}%
\left( V_{G}-V_{G}^{NN-1}\right) ,  \label{eq:VthRNN-1} \\
V_{L}^{N+1N} &\equiv &+%
{\displaystyle{C_{G} \over C-C_{L}}}%
\left( V_{G}^{N+1N}-V_{G}\right) ,  \label{eq:VthLN+1N} \\
V_{R}^{N+1N} &\equiv &-%
{\displaystyle{C_{G} \over C_{L}}}%
\left( V_{G}^{N+1N}-V_{G}\right) .  \label{eq:VthRN+1N}
\end{eqnarray}
\end{mathletters}%
%
The region where the $N$ electron ground state is stable, the $N$th {\em %
Coulomb diamond}, is thus delimited by $V_{L}^{NN-1}\left( V_{G}\right)
,V_{R}^{N+1N}\left( V_{G}\right) <V<V_{R}^{NN-1}\left( V_{G}\right)
,V_{L}^{N+1N}\left( V_{G}\right) $ for gate voltages $%
V_{G}^{NN-1}<V_{G}<V_{G}^{N+1N}$. The maximal bias voltage that can be
applied without lifting the Coulomb blockade is 
\begin{equation}
V_{\text{max}}^{N}=\frac{C_{G}}{C}\left( V_{G}^{N+1N}-V_{G}^{NN-1}\right)
=\delta +U.  \label{eq:VthmaxN}
\end{equation}
This is achieved at the intersection of the threshold lines (\ref
{eq:VthRNN-1}) and (\ref{eq:VthLN+1N})\ (the top of the diamond) at gate
voltage $V_{G\text{ max}}^{N}=V_{G}^{NN-1}+\left( C_{L}/C\right)
V_{G}^{N+1N}-V_{G}^{NN-1})$. Similarly, $V=-V_{\text{max}}^{N}$ at the
intersection of lines (\ref{eq:VthLNN-1}) and (\ref{eq:VthRN+1N}) at $%
V_{G}=V_{G}^{N+1N}-\left( C_{L}/C\right) (V_{G}^{N+1N}-V_{G}^{NN-1})$. The
addition energies relative to the electro-chemical potentials can now be
expressed in terms of distance of the bias voltage to the (gate-voltage
dependent) thresholds value%
\begin{mathletters}%
%
\label{eq:add_en} 
\begin{eqnarray}
E_{NN-1} &=&\mu _{R}-\frac{C_{L}}{C}\left( V_{R}^{NN-1}\left( V_{G}\right)
-V\right) <\mu _{R}, \\
E_{N+1N} &=&\mu _{L}+\left( 1-\frac{C_{L}}{C}\right) \left(
V_{L}^{N+1N}\left( V_{G}\right) -V\right) >\mu _{L}.
\end{eqnarray}
\end{mathletters}%
%

Let us now consider the transport conditions for the $N$ electron {\em %
excited} state. We assume that the lowest lying excited state is a spin {\em %
triplet~}\cite{bib:Kondo}, 
\begin{eqnarray*}
|N_{1,+1}^{\prime }\rangle &=&a_{d^{\prime }\uparrow }^{\dagger
}a_{d\uparrow }^{\dagger }|0\rangle , \\
|N_{1,0}^{\prime }\rangle &=&\frac{1}{\sqrt{2}}\left( a_{d^{\prime }\uparrow
}^{\dagger }a_{d\downarrow }^{\dagger }+a_{d^{\prime }\downarrow }^{\dagger
}a_{d\uparrow }^{\dagger }\right) |0\rangle , \\
|N_{1,-1}^{\prime }\rangle &=&a_{d^{\prime }\downarrow }^{\dagger
}a_{d\downarrow }^{\dagger }|0\rangle .
\end{eqnarray*}
with energy $E_{N^{\prime }}=\Delta +E_{N}$. The prime indicates the excited
state and the excitation energy $\Delta =\delta -J$ is just the spacing
between the two active levels $d$ and $d^{\prime }$ reduced by the exchange
energy. We concentrate on the change in the current due to this excited
state. The effect of excited states with higher energy (such as the spin
singlet at $E_{N}+\delta $) is disregarded. In order to have transport by
sequential tunneling, the transitions between the $N+1$ and $N-1$ electron
ground states and the $N$ electron excited state must be possible:%
\begin{mathletters}%
%
\label{eq:window'} 
\begin{eqnarray}
\mu _{L} &\geq &E_{N^{\prime }N-1}\geq \mu _{R},  \label{eq:window'_NN-1} \\
\mu _{L} &\geq &E_{N+1N^{\prime }}\geq \mu _{R}.  \label{eq:window'_N+1N}
\end{eqnarray}
\end{mathletters}%
%
where 
\begin{equation}
E_{N^{\prime }N-1}=E_{NN-1}+\Delta ,\quad E_{N+1N^{\prime }}=E_{N+1N}-\Delta
\label{eq:add_en'}
\end{equation}
The regions in the stability diagram where the conditions (\ref
{eq:window'_NN-1}) and (\ref{eq:window_N+1N}) are satisfied, are delimited
by another four thresholds lines for the bias voltage%
\begin{mathletters}%
%
\label{eq:Vth'}: 
\begin{eqnarray}
V_{\alpha }^{N^{\prime }N-1}\left( V_{G}\right) &=&V_{\alpha }^{NN-1}\left(
V_{G}+\frac{C}{C_{G}}\Delta \right) , \\
V_{\alpha }^{N^{\prime }N+1}\left( V_{G}\right) &=&V_{\alpha }^{NN+1}\left(
V_{G}-\frac{C}{C_{G}}\Delta \right) ,
\end{eqnarray}
\end{mathletters}%
%
respectively, where $\alpha =L,R$. These are simply the regions in Fig.~\ref
{fig:spectr} where (\ref{eq:window'_NN-1}) and (\ref{eq:window_N+1N}) are
satisfied, shifted in gate voltage by $\left( C/C_{G}\right) \Delta $
towards the center of the Coulomb diamond. Outside the Coulomb diamond the
sequential tunneling transitions to and from the $N$ electron excited state
gives rise to an increase in the current. By measuring the differential
conductance a spectroscopy of excited states can be performed.~\cite
{bib:excpeaks} {\em Within} the Coulomb diamond the $N$ electron ground
state is stable with respect to sequential tunneling. If the tunnel coupling
to the leads is weak, the transport is suppressed within the Coulomb
diamond. As the tunnel coupling is enhanced transport becomes possible as a
result of higher order tunneling processes. At low bias voltage, $V<\Delta $%
, only {\em elastic co-tunneling} is allowed by energy conservation. Charge
is transported through the dot without exciting it. At higher bias voltage, $%
V\geq \Delta $, {\em inelastic co-tunneling} processes excite the dot and
give rise to an additional current. Since these processes transfer energy
between the dot and the leads, the co-tunneling is called inelastic.~\cite
{bib:cot1} The dot switches between the $N$ electron ground and excited
state and in certain cases the $N\pm 1$ electron ground states can be
occupied, too. These two regimes are considered separately in the next two
sections. We point out that the model introduced here can be applied to a
limited range of bias and gate voltages in an experimental situation where
the capacitances and single particle energies provide a sufficiently
accurate parametrization of the many-body spectrum of the dot.

\section{Elastic co-tunneling}

\label{sec:el}At low bias voltage $V<\Delta $, a tunneling process of second
order in the transition amplitude cannot populate the excited state.
However, in this order of tunneling, transport is possible via elastic
co-tunneling without exciting the dot. This is the dominant contribution to
the current at sufficiently low temperatures and voltages. We do not
consider here the integer-spin Kondo physics on a small energy scale $T_{K}$%
, which has recently been observed:$~$\cite{bib:Kondo,bib:Kondo2} we assume $%
T,V\gg T_{K}$. It was pointed out in Refs.~ \cite{bib:Glattli,bib:KangMin}
that elastic co-tunneling is the only process limiting the operation of a
quantum dot as a single electron transistor when the level quantization $%
\delta $ is comparable with the charging energy $U$.

In a co-tunneling process two tunneling events occur in a very short time
interval. For instance, an electron is first injected into the dot from the
left lead and then emitted to the right lead. On this small time scale the
two tunneling events are coherent. The energy of the total system (dot and
leads) is sufficiently uncertain for the process not to be completely
blocked. However, the transition amplitude for this process is inversely
proportional to the energy difference between the initial and intermediate
state of the total system.~\cite{bib:FeynmanHibbs} The amplitude for
transferring an electron in the opposite sequence (first emit an electron,
then inject one) via another intermediate state must also be added if the
final state of the system is the same. From the total second order amplitude
(or matrix element) one obtains the tunneling rate by statistically
averaging over all initial and final states of the leads.~\cite{bib:cot1}
Thus a current can pass through the dot although the $N-1$ and $N+1$
electron states are only virtually accessible.

Let us describe the calculation of the tunneling rate for this simple case
in order to skip the details of similar calculations below. Consider the
initial state of the system where two electrons are in level $l$ of the left
leads and the dot in the ground state, denoted by $|N_{0,0}\rangle
a_{l\uparrow }^{\dagger }a_{l\downarrow }^{\dagger }|\rangle $ where $%
|\rangle $ denotes the state of the other electrons in the noninteracting
leads. Applying the tunneling operators $H_{TLd}^{\dagger }$ and $H_{RLd}$
in the two possible sequences, one obtains the final state $|N_{0,0}\rangle 
\hat{F}|\rangle $ where the operator $\hat{F}$ is given in Table~\ref
{tab:trans}. The amplitude for this process is written as $\sqrt{2}%
M_{N\leftarrow N}$ where the factor $\sqrt{2}$ is due to the spin degeneracy
of the levels in the left lead. The explicit expression for $M_{N\leftarrow
N}$ is given in Table~\ref{tab:ampl}. Introducing the tunneling rate 
\begin{equation}
\Gamma _{N\leftarrow N}=2\pi \sum_{l,r}\left| M_{N\leftarrow N}\right|
^{2}f\left( \varepsilon _{l}-\mu _{L}\right) \left[ 1-f\left( \varepsilon
_{r}-\mu _{R}\right) \right] \delta \left( \varepsilon _{l}-\varepsilon
_{r}\right) ,  \label{eq:G[NN]def}
\end{equation}
we write for the elastic co-tunneling current 
\begin{equation}
I=2\Gamma _{N\leftarrow N}.  \label{eq:Iel}
\end{equation}
The sum over all possible states of the leads with the Fermi distribution $%
f\left( \varepsilon \right) =1/\left( 1+e^{\varepsilon /T}\right) $ can be
converted into an integration and gives 
\begin{eqnarray}
\Gamma _{N\leftarrow N} &=&\frac{1}{2\pi }\left[ \Gamma _{dd}^{L}\Gamma
_{dd}^{R}\tau \left( E_{NN-1},V\right) +\Gamma _{d^{\prime }d^{\prime
}}^{L}\Gamma _{d^{\prime }d^{\prime }}^{R}\tau \left( E_{N+1N},V\right)
\right.  \nonumber \\
&&+\left. \Gamma _{dd^{\prime }}^{L}\Gamma _{d^{\prime }d}^{R}\tau ^{\prime
}\left( E_{NN-1},E_{N+1N},V\right) \right]  \label{eq:G[NN]}
\end{eqnarray}
Here we introduced the rates $\Gamma _{dd}^{L}=2\pi \sum_{l}\left|
t_{ld}\right| ^{2}\delta \left( \varepsilon -\varepsilon _{l}\right) $ and $%
\Gamma _{d^{\prime }d^{\prime }}^{L}$ ($d\rightarrow d^{\prime }$) for
sequential tunneling into level $d$ and $d^{\prime }$ and we assumed that
these depend only weakly on the energy $\varepsilon $. The same applies to
the quantity $\Gamma _{dd^{\prime }}^{L}=2\pi \sum_{l}t_{ld}^{\ast
}t_{ld^{\prime }}\delta \left( \varepsilon -\varepsilon _{l}\right) $ in the
last term on the right-hand side of Eq. (\ref{eq:G[NN]}), which takes into
account the interference between the two possible sequences for the
tunneling events (``paths'' in energy space); $\Gamma _{dd^{\prime
}}^{R},\Gamma _{dd}^{R}$,$\Gamma _{d^{\prime }d^{\prime }}^{R}$ are defined
similarly ($L,l\rightarrow R,r$). All these quantities are treated as
parameters here. In the expression for the rate (\ref{eq:G[NN]}) the
explicit ($V$) and implicit dependence [through the addition energies; cf.
Eqs. (\ref{eq:add_en})] on the voltages is incorporated in the functions
(explicitly evaluated in the limit $T\rightarrow 0$)%
\begin{mathletters}%
%
\begin{eqnarray}
\tau \left( E,V\right) &=&\int_{-\infty }^{\infty }d\varepsilon \frac{%
f\left( \varepsilon -\mu _{L}\right) \left[ 1-f\left( \varepsilon -\mu
_{R}\right) \right] }{\left( \varepsilon -E\right) ^{2}}  \label{eq:tau} \\
&=&\frac{V}{\left( \mu _{L}-E\right) \left( \mu _{R}-E\right) }\Theta \left(
V\right) ,  \nonumber \\
\tau ^{\prime }\left( E,E^{\prime },V\right) &=&\int_{-\infty }^{\infty
}d\varepsilon \frac{f\left( \varepsilon -\mu _{L}\right) \left[ 1-f\left(
\varepsilon -\mu _{R}\right) \right] }{\left| \varepsilon -E\right| \left|
\varepsilon -E^{\prime }\right| }  \label{eq:tau'} \\
&=&\frac{1}{\left| E^{\prime }-E\right| }\left( \left| \ln \frac{\mu _{L}-E}{%
\mu _{R}-E}\right| +\left| \ln \frac{\mu _{L}-E^{\prime }}{\mu
_{R}-E^{\prime }}\right| \right) \Theta \left( V\right) ,  \nonumber
\end{eqnarray}
\end{mathletters}%
%
Here $\Theta $ is the unit step function and $\lim_{E^{\prime }\rightarrow
E}\tau ^{\prime }\left( E,E^{\prime },V\right) =\tau \left( E,V\right) $.
The voltage dependence will be important below. The rate (\ref{eq:G[NN]})\
diverges near the edges of the Coulomb diamond where sequential tunneling
becomes dominant [cf. condition (\ref{eq:window})]. Here the second order
perturbation theory for the amplitudes breaks down. However, well within the
Coulomb diamond we have $\Gamma _{N\leftarrow N}\ll \Gamma _{dd^{\prime
}}^{L,R},\Gamma _{dd}^{L,R},\Gamma _{d^{\prime }d^{\prime }}^{L,R}$ and
perturbation theory can be used. This is the regime of interest here.

\section{Inelastic co-tunneling}

\label{sec:inel}If the bias voltage can supply the necessary excitation
energy, $V\geq \Delta $, then the excited state can be populated by
inelastic co-tunneling events.~\cite{bib:cot1} In such a coherent process,
an electron is transferred from the left to the right lead, thereby loosing
an amount $\Delta $ of its excess energy, which is supplied by the transport
voltage; this energy is transferred to the dot, which is thereby excited.
The excited state can relax to the $N$ electron ground state by two sorts of
processes depending on $\Delta $. If $\Delta $ is sufficiently small then
there is an energy barrier for injecting or emitting electrons [i.e.
conditions (\ref{eq:window'_NN-1}) and (\ref{eq:window'_N+1N}) are not
satisfied]. Thus the excited state is also Coulomb blockaded. The dot can
relax by three different inelastic co-tunneling processes. An electron can
be transferred from the left to the right lead again, but now gain an energy 
$\Delta $, thereby relaxing the dot. Similarly, an electron injected from
either lead can be excited above the Fermi level of that same lead. The dot
relaxes without a net charge transfer. This is the situation in the region
labeled (I) in Fig.~\ref{fig:spectr}. For the current the elastic
co-tunneling through the excited state is also important. If $\Delta $ is
sufficiently large, however, the excited state can decay to the ground state
by two sequential tunneling processes. There are three possibilities. Either
the dot can relax to the $N-1$ electron ground state and then return to the $%
N$ electron ground state [region (II) in Fig.~\ref{fig:spectr}]; or the dot
can be excited to the $N+1$ electron ground state from which it returns to
the $N$ electron ground state [region (II')]; or both processes can occur
[region (III)]. For positive bias voltage these regions are delimited by the
bias voltage threshold lines [Eqs. (\ref{eq:Vth}) and (\ref{eq:Vth'})] as
follows: 
\[
\begin{tabular}{clccccl}
$\text{(I)}$ & $\Delta $ & $<$ & $V$ & $<$ & $V_{L}^{N^{\prime
}N-1},V_{L}^{N+1N^{\prime }}$ & $\Delta <\frac{1}{3}V_{\text{max}}^{N}$ \\ 
$\text{(II)}$ & $\Delta ,V_{R}^{N^{\prime }N-1}$ & $<$ & $V$ & $<$ & $%
V_{R}^{NN-1},V_{L}^{N+1N^{\prime }}$ & $\Delta <\frac{1}{2}V_{\text{max}%
}^{N} $ \\ 
$\text{(II')}$ & $\Delta ,V_{L}^{N+1N^{\prime }}$ & $<$ & $V$ & $<$ & $%
V_{L}^{N+1N},V_{L}^{N^{\prime }N-1}$ & $\Delta <\frac{1}{2}V_{\text{max}%
}^{N} $ \\ 
(III) & $\Delta ,V_{R}^{N^{\prime }N-1},V_{L}^{N+1N^{\prime }}$ & $<$ & $V$
& $<$ & $V_{R}^{NN-1},V_{L}^{N+1N}$ & $\Delta <V_{\text{max}}^{N}$%
\end{tabular}
\]
Depending on the excitation energy $\Delta =\delta -J$ relative to the
maximum bias voltage $V_{\text{max}}^{N}=\delta +U$ some of these regions do
not exist. The different limiting cases are depicted in Fig.~\ref
{fig:spectrcases}. When the excitation energy is sufficiently small, $\Delta
<V_{\text{max}}^{N}/3$, all these regions exist. From this we find that the
Coulomb and/or exchange interaction should large relative to the level
spacing: $\delta <\frac{1}{2}U+\frac{3}{2}J$. Region (I) ceases to exist for
larger values of $\Delta $ (or $\delta $) while regions (II) and (II')
continue to exists as long as $\Delta <V_{\text{max}}^{N}/2$, i.e., $\delta
<U+2J$. Region (III) always exists since $\Delta <V_{\text{max}}^{N}$ is
satisfied for $U+J>0$. The situation is clearly more complicated than at low
voltage $V<\Delta $. The transport in each region needs to be considered
separately. Below we first consider the simplest region (I), where only the
ground and excited state participate in the transport. Then we extend this
analysis to the other regions.

\subsection{Region I - relaxation by co-tunneling}

This region exists only when the excited state lies sufficiently close to
the ground state. There are four states involved in the transport, of which
three are degenerate. In order to calculate the current we need the
nonequilibrium occupation of the states in the dot. We assume that at the
bias voltage where inelastic co-tunneling processes become important, $V\sim
\Delta $, the lifetime of the states in the dots is much smaller than the
typical time scale for building up Kondo-type correlations between the leads
and dots system. In terms of energies we thus assume $T_{K}\ll \Delta <V\ll
\Gamma $ where $T_{K}$ is the energy scale for such correlations. In this
limit we can use the following master equations~\cite{bib:mastereq1} to
describe the occupation of the ground state $\rho _{N_{0,0}}$ and of the
three triplet excited states $\rho _{N_{1,m}^{\prime }},m=0,\pm 1$:

\begin{mathletters}%
%
\label{eq:mastereq_I_spin} 
\begin{eqnarray}
\partial _{t}\rho _{N_{0,0}} &=&\sum_{i=1,2,3}\Gamma _{N\leftarrow N^{\prime
}}^{\left( i\right) }\sum_{m=0,\pm 1}\rho _{N_{1,m}^{\prime }}-3\Gamma
_{N^{\prime }\leftarrow N}\rho _{N_{0,0}}, \\
\partial _{t}\rho _{N_{1,0}^{\prime }} &=&\Gamma _{N^{\prime }\leftarrow
N}\rho _{N_{0,0}}+\frac{1}{2}\Gamma _{N^{\prime }\leftarrow N^{\prime
}}^{\left( 1\right) }\sum_{m=\pm 1}\rho _{N_{1,m}^{\prime }}-\left[ \Gamma
_{N^{\prime }\leftarrow N^{\prime }}^{\left( 1\right) }+\sum_{i=1,2,3}\Gamma
_{N\leftarrow N^{\prime }}^{\left( i\right) }\right] \rho _{N_{1,0}^{\prime
}},  \nonumber \\
&& \\
\partial _{t}\rho _{N_{1,m}^{\prime }} &=&\Gamma _{N^{\prime }\leftarrow
N}\rho _{N_{0,0}}+\frac{1}{2}\Gamma _{N^{\prime }\leftarrow N^{\prime
}}^{\left( 1\right) }\rho _{N_{1,0}^{\prime }}-\left[ \frac{1}{2}\Gamma
_{N^{\prime }\leftarrow N^{\prime }}^{\left( 1\right) }+\sum_{i=1,2,3}\Gamma
_{N\leftarrow N^{\prime }}^{\left( i\right) }\right] \rho _{N_{1,m}^{\prime
}},  \nonumber \\
&&
\end{eqnarray}
\end{mathletters}%
%
where $m=\pm 1$ and $\rho _{N_{0,0}}+\sum_{m=0,\pm 1}\rho _{N_{1,m}^{\prime
}}=1$. The current reads 
\begin{equation}
I=\left( 2\Gamma _{N\leftarrow N}+3\Gamma _{N^{\prime }\leftarrow N}\right)
\rho _{N_{0,0}}+\left( \frac{1}{2}\Gamma _{N^{\prime }\leftarrow N^{\prime
}}^{\left( 1\right) }+\Gamma _{N^{\prime }\leftarrow N^{\prime }}^{\left(
2\right) }+\Gamma _{N\leftarrow N^{\prime }}^{\left( 1\right) }\right)
\sum_{m=0,\pm 1}\rho _{N_{1,m}^{\prime }}.  \label{eq:Iinel_I_spin}
\end{equation}
The transitions between the states are schematically indicated in Fig.~\ref
{fig:trans}. The rates are calculated using the Golden Rule as was done for
Eq. (\ref{eq:G[NN]}). In Table~\ref{tab:trans} we have listed the final
state of the leads, the matrix element and the transition rate for each
transition. The explicit expressions for the matrix elements are given in
Table~\ref{tab:ampl}. Let us now calculate these rates and discuss the
transitions that they describe starting with the inelastic co-tunneling
processes.

Each excited state of the triplet is populated by inelastic co-tunneling
transition from the ground state with the same rate $\Gamma _{N^{\prime
}\leftarrow N}$. An electron with energy $\varepsilon _{l}$ and spin $s=\pm
1/2$ is first injected into the dot from lead $L$, bringing the dot in the
intermediate state $|N+1\rangle _{1/2,s}$, and is subsequently emitted into
the lead $R$ at an energy $\varepsilon _{r}=\varepsilon _{l}-\Delta $. The
other sequence is also possible, in which case the intermediate state of the
dot is $|N-1\rangle _{1/2,-s}$. The sum of the amplitudes for these
processes yields the matrix element $M_{N^{\prime }\leftarrow N}$, listed in
Table~\ref{tab:ampl}. The inelastic co-tunneling rate for exciting the dot
is then reads [cf. Eqs. (\ref{eq:tau}) and (\ref{eq:tau'})]%
\begin{mathletters}%
%
\begin{eqnarray}
\Gamma _{N^{\prime }\leftarrow N} &=&2\pi \sum_{l,r}\left| M_{N^{\prime
}\leftarrow N}\right| ^{2}f\left( \varepsilon _{l}-\mu _{L}\right) \left[
1-f\left( \varepsilon _{r}-\mu _{R}\right) \right] \delta \left( \varepsilon
_{l}-\varepsilon _{r}-\Delta \right)  \label{eq:G[N'N']def} \\
&=&\frac{1}{2\pi }\Gamma _{d^{\prime }d^{\prime }}^{L}\Gamma _{dd}^{R}\left[
\tau \left( E_{NN-1},V-\Delta \right) +\tau \left( E_{N+1N}-\Delta ,V-\Delta
\right) \right. \quad \quad \quad  \nonumber \\
&&\left. +2\tau ^{\prime }\left( E_{NN-1},E_{N+1N}-\Delta ,V-\Delta \right) 
\right]  \label{eq:G[N'N']}
\end{eqnarray}
\end{mathletters}%
%
Conversely, each state of the triplet can relax to the singlet ground state
by inelastic co-tunneling. The rate for these processes must be calculated
separately because they result in a different final state of the leads. The
matrix element $M_{N\leftarrow N^{\prime }}^{\left( 1\right) }$ in Table~\ref
{tab:ampl} describes the transfer of an electron from the left to the right
lead starting in any of the three triplet sublevels of the excited state.
The energy of the emitted electron is increased by the excitation energy $%
\varepsilon _{r}=\varepsilon _{l}+\Delta $. The inelastic co-tunneling rate
for this relaxation process is [cf. Eq. (\ref{eq:add_en'})]%
\begin{mathletters}%
%
\begin{eqnarray}
\Gamma _{N\leftarrow N^{\prime }}^{\left( 1\right) } &=&2\pi
\sum_{l,r}\left| M_{N\leftarrow N^{\prime }}^{\left( 1\right) }\right|
^{2}f\left( \varepsilon _{l}-\mu _{L}\right) \left[ 1-f\left( \varepsilon
_{r}-\mu _{R}\right) \right] \delta \left( \varepsilon _{l}+\Delta
-\varepsilon _{r}\right)  \label{eq:G[NN'](1)def} \\
&=&\frac{1}{2\pi }\Gamma _{d^{\prime }d^{\prime }}^{L}\Gamma _{dd}^{R}\left[
\tau \left( E_{N^{\prime }N-1},V+\Delta \right) +\tau \left( E_{N+1N^{\prime
}}+\Delta ,V+\Delta \right) \right. \quad \quad \quad  \nonumber \\
&&\left. +2\tau ^{\prime }\left( E_{N^{\prime }N-1},E_{N+1N^{\prime
}}+\Delta ,V+\Delta \right) \right] .  \label{eq:G[NN'](1)}
\end{eqnarray}
\end{mathletters}%
%
This rate explicitly depends on the voltage because the voltage window (both 
$\mu _{L}$ and $\mu _{R}$ appear in the Fermi functions). The second process
involves the injection of an electron at energy $\varepsilon _{l}$ from the
left lead and the emission of an electron to an empty state in the same lead
at a higher energy $\varepsilon _{l^{\prime }}=\varepsilon _{l}+\Delta .$
The matrix element $M_{N\leftarrow N^{\prime }}^{\left( 2\right) }$ is thus
found by replacing $r\rightarrow l^{\prime }$, cf. Table~\ref{tab:ampl}. The
tunneling rate is similarly found by replacing $r,R\rightarrow l^{\prime },L$
in Eq. (\ref{eq:G[NN'](1)def}) and from the result (\ref{eq:G[NN'](1)}) we
obtain: 
\[
\Gamma _{N\leftarrow N^{\prime }}^{\left( 2\right) }=\frac{\Gamma _{dd}^{L}}{%
\Gamma _{dd}^{R}}\Gamma _{N\leftarrow N^{\prime }}^{\left( 1\right) }\left(
V=0\right) . 
\]
The third process involves the injection and emission of an electron to and
from the right lead: replacing $L\rightarrow R$ we find: 
\[
\Gamma _{N\leftarrow N^{\prime }}^{\left( 3\right) }=\frac{\Gamma
_{d^{\prime }d^{\prime }}^{R}}{\Gamma _{d^{\prime }d^{\prime }}^{L}}\Gamma
_{N\leftarrow N^{\prime }}^{\left( 1\right) }\left( V=0\right) . 
\]

Let us now consider the elastic co-tunneling processes in which electrons
are transferred from the left to the right lead. These involve four
spin-degenerate intermediate states of the dot. In addition to the four
ground states (\ref{eq:gsN-1N+1}), there are four other intermediate {\em %
excited} states%
\begin{mathletters}%
%
\begin{eqnarray*}
|\left( N-1\right) _{1/2,+1/2}^{\prime }\rangle &=&a_{d^{\prime }\uparrow
}^{\dagger }|0\rangle ,\quad |\left( N+1\right) _{1/2,+1/2}^{\prime }\rangle
=a_{d^{\prime }\uparrow }^{\dagger }a_{d^{\prime }\downarrow }^{\dagger
}a_{d\uparrow }^{\dagger }|0\rangle , \\
|\left( N-1\right) _{1/2,-1/2}^{\prime }\rangle &=&a_{d^{\prime }\downarrow
}^{\dagger }|0\rangle ,\quad |\left( N+1\right) _{1/2,-1/2}^{\prime }\rangle
=a_{d^{\prime }\uparrow }^{\dagger }a_{d^{\prime }\downarrow }^{\dagger
}a_{d\downarrow }^{\dagger }|0\rangle .
\end{eqnarray*}
\end{mathletters}%
%
The addition energies for the transitions from the $N$ electron excited
state to these states are%
\begin{mathletters}%
%
\begin{eqnarray*}
E_{N^{\prime }N-1^{\prime }} &=&E_{NN-1}+\Delta -\delta =E_{NN-1}+J, \\
E_{N+1^{\prime }N^{\prime }} &=&E_{N+1N}+\delta -\Delta =E_{N+1N}-J.
\end{eqnarray*}
\end{mathletters}%
%
These energies lie closer to the Fermi energies than $E_{NN-1}$ and $%
E_{N+1N} $. Therefore the contributions of these intermediate states to
elastic co-tunneling amplitudes are larger than those in the amplitude $%
M_{N\leftarrow N}$ and they cannot be disregarded. Three types of elastic
co-tunneling processes must be considered separately. Two of these processes
do not involve a flip of the spin of the transmitted electron. The two
different final states of the leads as indicated in Table~\ref{tab:trans}.
The first type of process occurs with the same rate $\frac{1}{2}\Gamma
_{N^{\prime }\leftarrow N^{\prime }}^{\left( 1\right) }$ for all states of
the triplet. The second type is only possible for the two states with
nonzero spin projection and occurs with a rate $\frac{1}{2}\Gamma
_{N^{\prime }\leftarrow N^{\prime }}^{\left( 2\right) }$. The rates are
calculated from the matrix elements $M_{N^{\prime }\leftarrow N^{\prime
}}^{\left( 1,2\right) }$ in Table~\ref{tab:ampl} and give [compare this with
Eq. (\ref{eq:G[NN]})]%
\begin{mathletters}%
%
\begin{eqnarray}
\Gamma _{N^{\prime }\leftarrow N^{\prime }}^{\left( 1,2\right) } &=&2\pi
\sum_{l,r}\left| M_{N^{\prime }\leftarrow N^{\prime }}^{\left( 1,2\right)
}\right| ^{2}f\left( \varepsilon _{l}-\mu _{L}\right) \left[ 1-f\left(
\varepsilon _{r}-\mu _{R}\right) \right] \delta \left( \varepsilon
_{l}-\varepsilon _{r}\right)  \label{eq:G[N',N'](1,2)def} \\
&=&\frac{1}{2\pi }\left\{ \Gamma _{dd}^{L}\Gamma _{dd}^{R}\left[ \tau \left(
E_{N^{\prime }N-1},V\right) +\tau \left( E_{N+1N^{\prime }},V\right) \pm
2\tau ^{\prime }\left( E_{N^{\prime }N-1},E_{N+1N^{\prime }},V\right) \right]
\right.  \nonumber \\
&&+\Gamma _{d^{\prime }d^{\prime }}^{L}\Gamma _{d^{\prime }d^{\prime }}^{R} 
\left[ \tau \left( E_{N^{\prime }N-1^{\prime }},V\right) +\tau \left(
E_{N+1^{\prime }N^{\prime }},V\right) \pm 2\tau ^{\prime }\left(
E_{N^{\prime }N-1^{\prime }},E_{N+1^{\prime }N^{\prime }},V\right) \right] 
\nonumber \\
&&+2%
\mathop{\rm Re}%
\Gamma _{dd^{\prime }}^{L}\Gamma _{d^{\prime }d}^{R}\left[ \tau ^{\prime
}\left( E_{N^{\prime }N-1},E_{N^{\prime }N-1^{\prime }},V\right) \pm \tau
^{\prime }\left( E_{N^{\prime }N-1},E_{N+1^{\prime }N^{\prime }},V\right)
\right.  \nonumber \\
&&\left. \left. +\tau ^{\prime }\left( E_{N+1N^{\prime }},E_{N+1^{\prime
}N^{\prime }},V\right) \pm \tau ^{\prime }\left( E_{N^{\prime }N-1^{\prime
}},E_{N+1N^{\prime }},V\right) \right] \right\} .  \label{eq:G[N',N'](1,2)}
\end{eqnarray}
\end{mathletters}%
%
These processes do not induce transitions between the states [they don not
contribute to Eqs. (\ref{eq:mastereq_I_spin})], but they do contribute to
the transfer of electrons [the rate appears in Eq. (\ref{eq:Iinel_I_spin})].
The third type of elastic co-tunneling process, involves a flip of the spin
of the transmitted electron and induces transitions between the triplet
states states. The triplet state $|N_{1,0}^{\prime }\rangle $ can decay to
both $|N_{1,1}^{\prime }\rangle $ and $|N_{1,-1}^{\prime }\rangle $ with a
rate equal to $\frac{1}{2}\Gamma _{N^{\prime }\leftarrow N^{\prime
}}^{\left( 2\right) }$. The latter two states can return to the former state
with the same transition rate.

Because there is no Zeeman energy or spin-orbit coupling in our model, the
transition rates between the states of the triplet balance each other, cf.
Fig.~\ref{fig:trans}. Therefore we can introduce the total occupations of
the excited state $\rho _{N^{\prime }}\equiv \sum_{m=0,\pm 1}\rho
_{N_{1,m}^{\prime }}$summed over the spin projection and omit the spin
indices $\rho _{N}\equiv \rho _{N_{0,0}}$. By adding the equations for the
triplet states we obtain a closed set of master equations for the total
occupations:%
\begin{mathletters}%
%
\label{eq:mastereq_I_tot} 
\begin{eqnarray}
\partial _{t}\rho _{N} &=&\sum_{i=1,2,3}\Gamma _{N\leftarrow N^{\prime
}}^{\left( i\right) }\rho _{N^{\prime }}-3\Gamma _{N^{\prime }\leftarrow
N}\rho _{N}, \\
\partial _{t}\rho _{N^{\prime }} &=&3\Gamma _{N^{\prime }\leftarrow N}\rho
_{N}-\sum_{i=1,2,3}\Gamma _{N\leftarrow N^{\prime }}^{\left( i\right) }\rho
_{N^{\prime }},
\end{eqnarray}
\end{mathletters}%
%
with $\rho _{N}+\rho _{N^{\prime }}=1$. The elastic co-tunneling transitions
between the triplet states have precisely cancelled. However, these
processes do contribute to the current (\ref{eq:Iinel_I_spin}). The fact
that the triplet state with zero spin projection transfers twice as much
charge per unit time ($2\times \frac{1}{2}\Gamma _{N^{\prime }\leftarrow
N^{\prime }}^{\left( 2\right) }$) as the other states ($\frac{1}{2}\Gamma
_{N^{\prime }\leftarrow N^{\prime }}^{\left( 2\right) }$) is precisely
balanced by the spin-flip processes that can only occur for states with
nonzero spin projection ($\frac{1}{2}\Gamma _{N^{\prime }\leftarrow
N^{\prime }}^{\left( 2\right) }$). The elastic co-tunneling contribution of
each state of the triplet is thus proportional to the same rate (\ref
{eq:Iinel_I_spin}) and we can also express the current in total occupations 
\begin{equation}
I=\left( 2\Gamma _{N\leftarrow N}+3\Gamma _{N^{\prime }\leftarrow N}\right)
\rho _{N}+\left( \frac{1}{2}\Gamma _{N^{\prime }\leftarrow N^{\prime
}}^{\left( 1\right) }+\Gamma _{N^{\prime }\leftarrow N^{\prime }}^{\left(
2\right) }+\Gamma _{N\leftarrow N^{\prime }}^{\left( 1\right) }\right) \rho
_{N^{\prime }}.  \label{eq:Iinel_Itot}
\end{equation}
Note that the inelastic co-tunneling relaxation processes with rates $\Gamma
_{N\leftarrow N^{\prime }}^{\left( 2.3\right) }$ do not appear here since
they do not involve charge transfer. In the stationary limit $%
\lim_{t\rightarrow \infty }\partial _{t}\rho _{N^{\prime }},\partial
_{t}\rho _{N}=0$ equations (\ref{eq:mastereq_I_tot}) are readily solved. The
current can be written as 
\begin{equation}
I=\frac{\left( \frac{1}{2}\Gamma _{N^{\prime }\leftarrow N^{\prime
}}^{\left( 1\right) }+\Gamma _{N^{\prime }\leftarrow N^{\prime }}^{\left(
2\right) }-2\Gamma _{N\leftarrow N}\right) +2\Gamma _{N\leftarrow N^{\prime
}}^{\left( 1\right) }+\sum_{i=2,3}\Gamma _{N\leftarrow N^{\prime }}^{\left(
i\right) }}{\sum_{i=1,2,3}\Gamma _{N^{\prime }\leftarrow N}^{\left( i\right)
}+3\Gamma _{N^{\prime }\leftarrow N}}3\Gamma _{N^{\prime }\leftarrow
N}+2\Gamma _{N\leftarrow N}.  \label{eq:Iinel_I}
\end{equation}
For $V<\Delta $ the excitation rate $\Gamma _{N^{\prime }\leftarrow N}$
vanishes and we recover the elastic co-tunneling current \ref{eq:Iel}. In
Fig.~\ref{fig:Iinel_I} we have plotted the current, differential conductance
and the co-tunneling rates as a function of bias voltages for the case $%
\Delta <V_{\text{max}}^{N}/3$. In this limit region (I) covers most of the
Coulomb diamond for $V\geq \Delta $. Apart from a clear kink in the current
and a step in the differential conductance, a sharp peak is present at the
step edge. This is clearly due to a change of the population of the excited
state.

\subsection{Regions (II), (II') and (III)\ - relaxation by sequential
tunneling}

The remain three regions where sequential tunneling transitions from the
excited state to $N-1$ and/or $N+1$ electron ground state are possible are
now easily included. The elastic co-tunneling processes involving the
excited state must now be disregarded ( and their contributions to the
current). The rates for these processes diverge at the boundaries of regime
(I) signalling that perturbation theory breaks down. Instead we write a full
set of master equations similar to (\ref{eq:mastereq_I_spin}) without the
transitions between the triplet states and now including the occupations of
the spin sublevels of the $N-1$ and/or $N+1$ electron ground states.
Proceeding as before we arrive at the following closed set of master
equations for the total occupations, including $\rho _{N\pm 1}\equiv
\sum_{m=\pm 1/2}\rho _{\left( N\pm 1\right) _{1/2,m}}$:%
\begin{mathletters}%
%
\label{eq:mastereq_III_total} 
\begin{eqnarray}
\partial _{t}\rho _{N-1} &=&\alpha _{-}\Gamma _{d^{\prime }d^{\prime
}}^{R}\rho _{N^{\prime }}-\left( \alpha _{-}\Gamma _{dd}^{R}+\alpha
_{+}\Gamma _{d^{\prime }d^{\prime }}^{L}\right) \rho _{N-1}, \\
\partial _{t}\rho _{N+1} &=&\alpha _{+}\Gamma _{dd}^{L}\rho _{N^{\prime
}}-\left( \alpha _{-}\Gamma _{d^{\prime }d^{\prime }}^{L}+\alpha _{+}\Gamma
_{d^{\prime }d^{\prime }}^{R}\right) \rho _{N+1}, \\
\partial _{t}\rho _{N} &=&+\alpha _{-}\Gamma _{dd}^{L}\rho _{N-1}+\alpha
_{+}\Gamma _{d^{\prime }d^{\prime }}^{R}\rho _{N+1}-3\Gamma _{N^{\prime
}\leftarrow N}\rho _{N}, \\
\partial _{t}\rho _{N^{\prime }} &=&\alpha _{-}\Gamma _{d^{\prime }d^{\prime
}}^{L}\rho _{N-1}+\alpha _{+}\Gamma _{dd}^{R}\rho _{N+1}+3\Gamma _{N^{\prime
}\leftarrow N}\rho _{N}  \nonumber \\
&&-\left( +\alpha _{-}\Gamma _{d^{\prime }d^{\prime }}^{R}\rho _{N^{\prime
}}+\alpha _{+}\Gamma _{dd}^{L}\rho _{N^{\prime }}\right) \rho _{N^{\prime }},
\end{eqnarray}
\end{mathletters}%
%
The current reads 
\begin{equation}
I=\left( 2\Gamma _{N\leftarrow N}+3\Gamma _{N^{\prime }\leftarrow N}\right)
\rho _{N}+\alpha _{-}\Gamma _{d^{\prime }d^{\prime }}^{R}\rho _{N^{\prime
}}+\alpha _{+}\left( \Gamma _{dd}^{R}+\Gamma _{d^{\prime }d^{\prime
}}^{R}\right) \rho _{N+1}.  \label{eq:Iinel_III_tot}
\end{equation}
These expression covers regions (II) [$\alpha _{-}=1,\alpha _{+}=0$], (II') [%
$\alpha _{-}=0,\alpha _{+}=1$] and (III) [$\alpha _{\pm }=1$]. The solution
for all cases can be written in the form 
\begin{equation}
I=\frac{2K+K^{\prime }-2\kappa \Gamma _{N\leftarrow N}+2\Gamma _{N^{\prime
}\leftarrow N}^{\left( 1\right) }}{K+3\kappa \Gamma _{N^{\prime }\leftarrow
N}}3\Gamma _{N^{\prime }\leftarrow N}+2\Gamma _{N\leftarrow N}.
\label{eq:Iinel_III}
\end{equation}
The coefficients appearing here are constants that characterize the
differences between the three regions:%
\begin{mathletters}%
%
\begin{eqnarray*}
K &=&\alpha _{-}\Gamma _{d^{\prime }d^{\prime }}^{R}\frac{\Gamma _{dd}^{L}}{%
\Gamma _{dd}^{L}+\Gamma _{d^{\prime }d^{\prime }}^{L}}+\alpha _{+}\Gamma
_{d^{\prime }d^{\prime }}^{L}\frac{\Gamma _{dd}^{R}}{\Gamma _{dd}^{R}+\Gamma
_{d^{\prime }d^{\prime }}^{R}}, \\
K^{\prime } &=&\alpha _{-}\Gamma _{d^{\prime }d^{\prime }}^{R}\frac{\Gamma
_{d^{\prime }d^{\prime }}^{L}}{\Gamma _{dd}^{L}+\Gamma _{d^{\prime
}d^{\prime }}^{L}}+\alpha _{+}\Gamma _{d^{\prime }d^{\prime }}^{L}\frac{%
\Gamma _{d^{\prime }d^{\prime }}^{R}}{\Gamma _{dd}^{R}+\Gamma _{d^{\prime
}d^{\prime }}^{R}}, \\
\kappa  &=&\alpha _{-}\frac{\Gamma _{d^{\prime }d^{\prime }}^{R}}{\Gamma
_{dd}^{L}+\Gamma _{d^{\prime }d^{\prime }}^{L}}+\alpha _{+}\frac{\Gamma
_{d^{\prime }d^{\prime }}^{L}}{\Gamma _{dd}^{R}+\Gamma _{d^{\prime
}d^{\prime }}^{R}}+1.
\end{eqnarray*}
\end{mathletters}%
%
where $K+K^{\prime }=\alpha _{-}\Gamma _{d^{\prime }d^{\prime }}^{R}+\alpha
_{+}\Gamma _{d^{\prime }d^{\prime }}^{L}$. In Fig.~\ref{fig:Iinel_III} we
have plotted the co-tunneling rates, the occupations of the states and the
differential conductance as a function of the bias voltages for $\Gamma
_{dd}^{\alpha }=\Gamma _{d^{\prime }d^{\prime }}^{\alpha }=\Gamma ,\alpha
=L,R$ for a typical case where $V_{\text{max}}^{N}/2\leq \Delta $. In this
limit only region (III) exists. The qualitative result is the same as in
region (I). The essential condition for the peak in the conductance is that
the total rate for charge transport through the excited state is bigger than
the rate $\Gamma _{N^{\prime }\leftarrow N}$ at which this state is
populated.

\section{Conclusions}

\label{sec:concl1}We have analyzed the finite bias transport {\em within} a
Coulomb blockaded region of a quantum dot with an even number of electrons.
The differential conductance through the dot was found to exhibit a step
within the Coulomb diamond at a bias voltage equal to the excitation energy.
At the step edge an additional peak appears that is caused by a change in
the nonequilibrium populations of the dot states due the voltage dependence
of the co-tunneling rates. This may explain the recent observations in the
transport spectroscopy of semi-conductor quantum dots by De Franceschi et.
al and of carbon nanotubes by Nyg\aa rd et. al. As pointed out in Ref.~\cite
{bib:SCdot1} the sharp feature in the differential conductance can be used
to do a spectroscopy of excited states with increased resolution. We have
considered two well-separated regimes where the excited state is depopulated
by inelastic co-tunneling and sequential tunneling, respectively. It is
expected that in the crossover regime the results do not change
qualitatively since the inelastic co-tunneling processes that excite the dot
form the bottleneck for the transport. The rate for these processes as
calculated here is correct also in the crossover regime. The calculations
presented here can be improved by including (additive) contributions of more
virtual states of the dot (more than two orbitals) to the transition
amplitudes, which determine the co-tunneling rates. Also, the situation
where a second $N$-electron excited state lies the Coulomb diamond can be
treated in the same way as single excited state. The occupations of all
transport states can be calculated from an extended set of master equations.
One must now also include co-tunneling transitions between the two excited
states. The rates that enter in these equations have the same form as those
calculated here. Furthermore, the peak at the onset of inelastic
co-tunneling is broadened by thermal fluctuations and level broadening
effects, as pointed out in Ref.~\cite{bib:SCdot1}. The effect of a electron
finite temperature can be included in our calculations. Level broadening
effects require a nonperturbative scheme, which is beyond the scope of this
paper. It is clear, however, that near the onset of inelastic
co-tunneling, the broadening of the peak, which is determined the decay rate
of the excited state (either by sequential or inelastic co-tunneling), is
much larger than the rate for exciting the dot.

\section{Acknowledgements}

We would like to thank S. De Franceschi for bringing this problem to our
attention and for useful discussions. One of us (M.R.W) would like to thank
M. Eto and Y. Tokura for valuable discussions. This work is part of the
research program of the ``Stichting voor Fundamenteel Onderzoek der
Materie'' (FOM), which is financially supported by the ''Nederlandse
Organisatie voor Wetenschappelijk Onderzoek'' (NWO) and the NEDO project
NTDP-98.

\begin{table}[tbp]
\caption{Transitions from initial state $|i\rangle a_{l\uparrow }^{\dagger
}a_{l\downarrow }^{\dagger }|\rangle $ to final state $|f\rangle \hat{F}%
|\rangle $ with the second order matrix elements and corresponding rates.
The explicit expressions for the matrix elements are given in Table~\ref
{tab:ampl}. The rates are given in the text. Here $l$ and $r^{\prime }$
label occupied states in the left and right lead, respectively; $l^{\prime }$
and $r$ label unoccupied states in the left and right lead, respectively.}
\label{tab:trans}
\begin{tabular}{cllccc}
type & $i\rightarrow $ & $f$ & $\hat{F}$ & $\left| M_{Ff\leftarrow
iI}\right| $ & $\Gamma _{f\leftarrow i}$ \\ \hline\hline
elastic & \multicolumn{1}{|l}{$N_{0,0}$} & $N_{0,0}$ & $\frac{1}{\sqrt{2}}%
\left( a_{l\downarrow }^{\dagger }a_{r\uparrow }^{\dagger }-a_{l\uparrow
}^{\dagger }a_{r\downarrow }^{\dagger }\right) $ & $\sqrt{2}M_{N\leftarrow
N} $ & $2\Gamma _{N\leftarrow N}$ \\ \hline
inelastic & \multicolumn{1}{|l}{$N_{0,0}$} & $N_{1,0}^{\prime }$ & $\frac{1}{%
\sqrt{2}}\left( a_{l\downarrow }^{\dagger }a_{r\uparrow }^{\dagger
}+a_{l\uparrow }^{\dagger }a_{r\downarrow }^{\dagger }\right) $ & $%
M_{N^{\prime }\leftarrow N}$ & $\Gamma _{N^{\prime }\leftarrow N}$ \\ 
& \multicolumn{1}{|l}{$N_{0,0}$} & $N_{1,+1}^{\prime }$ & $a_{l\downarrow
}^{\dagger }a_{r\downarrow }^{\dagger }$ &  &  \\ 
& \multicolumn{1}{|l}{$N_{0,0}$} & $N_{1,-1}^{\prime }$ & $a_{l\uparrow
}^{\dagger }a_{r\uparrow }^{\dagger }$ &  &  \\ \hline
elastic & \multicolumn{1}{|l}{$N_{1,m}^{\prime }$} & $N_{1,m}^{\prime
},m=\pm 1,0$ & $\frac{1}{\sqrt{2}}\left( a_{l\downarrow }^{\dagger
}a_{r\uparrow }^{\dagger }-a_{l\uparrow }^{\dagger }a_{r\downarrow
}^{\dagger }\right) $ & $M_{N^{\prime }\leftarrow N^{\prime }}^{\left(
1\right) }/\sqrt{2}$ & $\frac{1}{2}\Gamma _{N^{\prime }\leftarrow N^{\prime
}}^{\left( 1\right) }$ \\ \cline{2-6}
& \multicolumn{1}{|l}{$N_{1,m}^{\prime }$} & $N_{1,m}^{\prime },m=\pm 1$ (!)
& $\frac{1}{\sqrt{2}}\left( a_{l\downarrow }^{\dagger }a_{r\uparrow
}^{\dagger }+a_{l\uparrow }^{\dagger }a_{r\downarrow }^{\dagger }\right) $ & 
$M_{N^{\prime }\leftarrow N^{\prime }}^{\left( 2\right) }/\sqrt{2}$ & $\frac{%
1}{2}\Gamma _{N^{\prime }\leftarrow N^{\prime }}^{\left( 2\right) }$ \\ 
\cline{2-6}
& \multicolumn{1}{|l}{$N_{1,0}^{\prime }$} & $N_{1,+1}^{\prime }$ & $%
a_{l\downarrow }^{\dagger }a_{r\downarrow }^{\dagger }$ & $M_{N^{\prime
}\leftarrow N^{\prime }}^{\left( 2\right) }/\sqrt{2}$ & $\frac{1}{2}\Gamma
_{N^{\prime }\leftarrow N^{\prime }}^{\left( 2\right) }$ \\ 
& \multicolumn{1}{|l}{$N_{1,0}^{\prime }$} & $N_{1,-1}^{\prime }$ & $%
a_{l\uparrow }^{\dagger }a_{r\uparrow }^{\dagger }$ &  &  \\ 
& \multicolumn{1}{|l}{$N_{1,+1}^{\prime }$} & $N_{1,0}^{\prime }$ & $%
a_{l\uparrow }^{\dagger }a_{r\uparrow }^{\dagger }$ &  &  \\ 
& \multicolumn{1}{|l}{$N_{1,-1}^{\prime }$} & $N_{1,0}^{\prime }$ & $%
a_{l\downarrow }^{\dagger }a_{r\downarrow }^{\dagger }$ &  &  \\ \hline
inelastic & \multicolumn{1}{|l}{$N_{1,0}^{\prime }$} & $N_{0,0}$ & $\frac{1}{%
\sqrt{2}}\left( a_{l\downarrow }^{\dagger }a_{r\uparrow }^{\dagger
}+a_{l\uparrow }^{\dagger }a_{r\downarrow }^{\dagger }\right) $ & $%
M_{N\leftarrow N^{\prime }}^{\left( 1\right) }$ & $\Gamma _{N\leftarrow
N^{\prime }}^{\left( 1\right) }$ \\ 
& \multicolumn{1}{|l}{$N_{1,+1}^{\prime }$} & $N_{0,0}$ & $a_{l\uparrow
}^{\dagger }a_{r\uparrow }^{\dagger }$ &  &  \\ 
& \multicolumn{1}{|l}{$N_{1,-1}^{\prime }$} & $N_{0,0}$ & $a_{l\downarrow
}^{\dagger }a_{r\downarrow }^{\dagger }$ &  &  \\ \cline{2-6}
& \multicolumn{1}{|l}{$N_{1,0}^{\prime }$} & $N_{0,0}$ & $\frac{1}{\sqrt{2}}%
\left( a_{l\downarrow }^{\dagger }a_{l^{\prime }\uparrow }^{\dagger
}+a_{l\uparrow }^{\dagger }a_{l^{\prime }\downarrow }^{\dagger }\right) $ & $%
M_{N\leftarrow N^{\prime }}^{\left( 2\right) }$ & $\Gamma _{N\leftarrow
N^{\prime }}^{\left( 2\right) }$ \\ 
& \multicolumn{1}{|l}{$N_{1,+1}^{\prime }$} & $N_{0,0}$ & $a_{l\uparrow
}^{\dagger }a_{l^{\prime }\uparrow }^{\dagger }$ &  &  \\ 
& \multicolumn{1}{|l}{$N_{1,-1}^{\prime }$} & $N_{0,0}$ & $a_{l\downarrow
}^{\dagger }a_{l^{\prime }\downarrow }^{\dagger }$ &  &  \\ \cline{2-6}
& \multicolumn{1}{|l}{$N_{1,0}^{\prime }$} & $N_{0,0}$ & $\frac{1}{\sqrt{2}}%
\left( a_{r^{\prime }\downarrow }^{\dagger }a_{r\uparrow }^{\dagger
}+a_{r^{\prime }\uparrow }^{\dagger }a_{r\downarrow }^{\dagger }\right) $ & $%
M_{N\leftarrow N^{\prime }}^{\left( 3\right) }$ & $\Gamma _{N\leftarrow
N^{\prime }}^{\left( 3\right) }$ \\ 
& \multicolumn{1}{|l}{$N_{1,+1}^{\prime }$} & $N_{0,0}$ & $a_{r^{\prime
}\uparrow }^{\dagger }a_{r\uparrow }^{\dagger }$ &  &  \\ 
& \multicolumn{1}{|l}{$N_{1,-1}^{\prime }$} & $N_{0,0}$ & $a_{r^{\prime
}\downarrow }^{\dagger }a_{r\downarrow }^{\dagger }$ &  & 
\end{tabular}
\end{table}

\begin{table}[tbp]
\caption{Second order matrix elements for the transitions listed in Table~%
\ref{tab:trans}.}
\label{tab:ampl}
\begin{tabular}{lccc}
\hline
$M_{N\leftarrow N}$ & $=$ & $%
{\displaystyle{t_{ld}^{\ast }t_{rd} \over \varepsilon _{r}-E_{NN-1}}}%
+%
{\displaystyle{t_{ld^{\prime }}^{\ast }t_{rd^{\prime }} \over E_{N+1N}-\varepsilon _{l}}}%
$ & elastic \\ 
$M_{N^{\prime }\leftarrow N^{\prime }}^{\left( 1,2\right) }$ & $=$ & $%
{\displaystyle{t_{ld}^{\ast }t_{rd} \over \varepsilon _{r}-E_{N^{\prime }N-1}}}%
+%
{\displaystyle{t_{ld^{\prime }}^{\ast }t_{rd^{\prime }} \over \varepsilon _{r}-E_{N^{\prime }N-1^{\prime }}}}%
$ &  \\ 
&  & $\pm \left( 
{\displaystyle{t_{ld}^{\ast }t_{rd} \over E_{N+1N^{\prime }}-\varepsilon _{l}}}%
+%
{\displaystyle{t_{ld^{\prime }}^{\ast }t_{rd^{\prime }} \over E_{N+1^{\prime }N^{\prime }}-\varepsilon _{l}}}%
\right) $ &  \\ \hline
$M_{N^{\prime }\leftarrow N}$ & $=$ & $%
{\displaystyle{t_{ld^{\prime }}^{\ast }t_{rd} \over \varepsilon _{r}-E_{NN-1}}}%
+%
{\displaystyle{t_{ld^{\prime }}^{\ast }t_{rd} \over E_{N+1N}-\varepsilon _{l}}}%
$ & inelastic \\ 
$M_{N\leftarrow N^{\prime }}^{\left( 1\right) }$ & $=$ & $%
{\displaystyle{t_{ld}^{\ast }t_{rd^{\prime }} \over \varepsilon _{r}-E_{N^{\prime }N-1}}}%
+%
{\displaystyle{t_{ld}^{\ast }t_{rd^{\prime }} \over E_{N+1N^{\prime }}-\varepsilon _{l}}}%
$ &  \\ 
$M_{N\leftarrow N^{\prime }}^{\left( 2\right) }$ & $=$ & $%
{\displaystyle{t_{ld}^{\ast }t_{l^{\prime }d^{\prime }} \over \varepsilon _{l^{\prime }}-E_{N^{\prime }N-1}}}%
+%
{\displaystyle{t_{ld}^{\ast }t_{l^{\prime }d^{\prime }} \over E_{N+1N^{\prime }}-\varepsilon _{l}}}%
$ &  \\ 
$M_{N\leftarrow N^{\prime }}^{\left( 3\right) }$ & $=$ & $%
{\displaystyle{t_{r^{\prime }d}^{\ast }t_{rd^{\prime }} \over \varepsilon _{r}-E_{N^{\prime }N-1}}}%
+%
{\displaystyle{t_{r^{\prime }d}^{\ast }t_{rd^{\prime }} \over E_{N+1N^{\prime }}-\varepsilon _{r^{\prime }}}}%
$ & 
\end{tabular}
\end{table}

\begin{figure}[tbp]
\caption{Stabilitity diagram of a quantum dot for the symmetric case $%
C_{L}/C_{G}=C_{R}/C_{G}\gg 1$. The skewed lines delimit the regions where
sequential tunneling is possible through the ground states [thick; Eqs. (\ref
{eq:Vth})] and the excited state [thin; Eqs. (\ref{eq:Vth'})]. The
horizontal thick lines mark the onset of inelastic co-tunneling.}
\label{fig:spectr}
\end{figure}

\begin{figure}[tbp]
\caption{Inelastic co-tunneling regions for the different cases of
excitation energy relative to the maximal bias voltage.}
\label{fig:spectrcases}
\end{figure}

\begin{figure}[tbp]
\caption{Transitions between the spin-singlet ground state and the
spin-triplet excited state. Elastic co-tunneling processes which do not
change the state (charge and spin projection) are indicated by a circle,
beginning and ending at the same state.}
\label{fig:trans}
\end{figure}

\begin{figure}[tbp]
\caption{Results for region (I) as a function of bias voltage in the center
of the Coulomb diamond: $\Delta /V_{\text{max}}^{N}=0.1$, $\protect\delta %
/U=5$, $U=1$, $\Gamma /U=0.05$, $C_{L}/C_{G}=C_{R}/C_{G}=0.1$. {\em Left}:
Co-tunneling rates relative to the sequential tunneling ratee, i.e., in
units of $\Gamma $. Elastic: (1) for the $N$ electron ground state $\Gamma
_{N\leftarrow N}$ (2) for the $N$ electron excited state $\Gamma _{N^{\prime
}\leftarrow N^{\prime }}^{\left( 1\right) }=\Gamma _{N^{\prime }\leftarrow
N^{\prime }}^{\left( 2\right) }$. Inelastic: (3) for exciting the dot $%
\Gamma _{N^{\prime }\leftarrow N}$ (4) for relaxing the dot with charge
transfer $\Gamma _{N\leftarrow N^{\prime }}^{\left( 1\right) }$ and (5)
without charge transfer $\Gamma _{N\leftarrow N^{\prime }}^{\left(
2,3\right) }$. {\em Middle: }Occupations of (1) the $N$ electron ground
state (2), the $N$ electron excited state and (3) the $N-1$ and $N+1$
electron ground states (coincide). {\em Right}: Differential conductance
calculated from expression (\ref{eq:Iinel_I}) for the current.}
\label{fig:Iinel_I}
\end{figure}

\begin{figure}[tbp]
\caption{Results in region (III) as a function of bias voltage in the center
of the Coulomb diamond: $\Delta /V_{\text{max}}^{N}=0.5$, $\protect\delta %
/U=5$ ,$U=1$,$\Gamma /U=0.04$, $C_{L}/C_{G}=C_{R}/C_{G}=0.1$. {\em Left}:
Co-tunneling rates relative to the sequential tunneling rate, i.e., in units
of $\Gamma $. Elastic: (1) for the $N$ elctron ground state $\Gamma
_{N\leftarrow N}$. Inelastic: (2) for exciting the dot $\Gamma _{N^{\prime
}\leftarrow N}$. {\em Middle}: Occupations of (1) the $N$ electron ground
state (2) the $N$ electron excited state and (3) the $N-1$ and $N+1$
electron ground states (coincide). {\em Right}: Differential conductance
calculated from expression (\ref{eq:Iinel_III}) for the current.}
\label{fig:Iinel_III}
\end{figure}

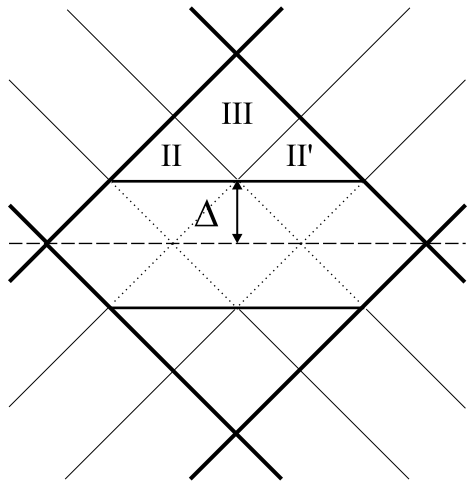
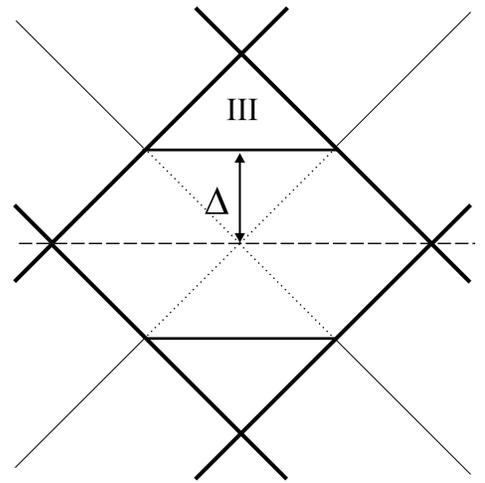
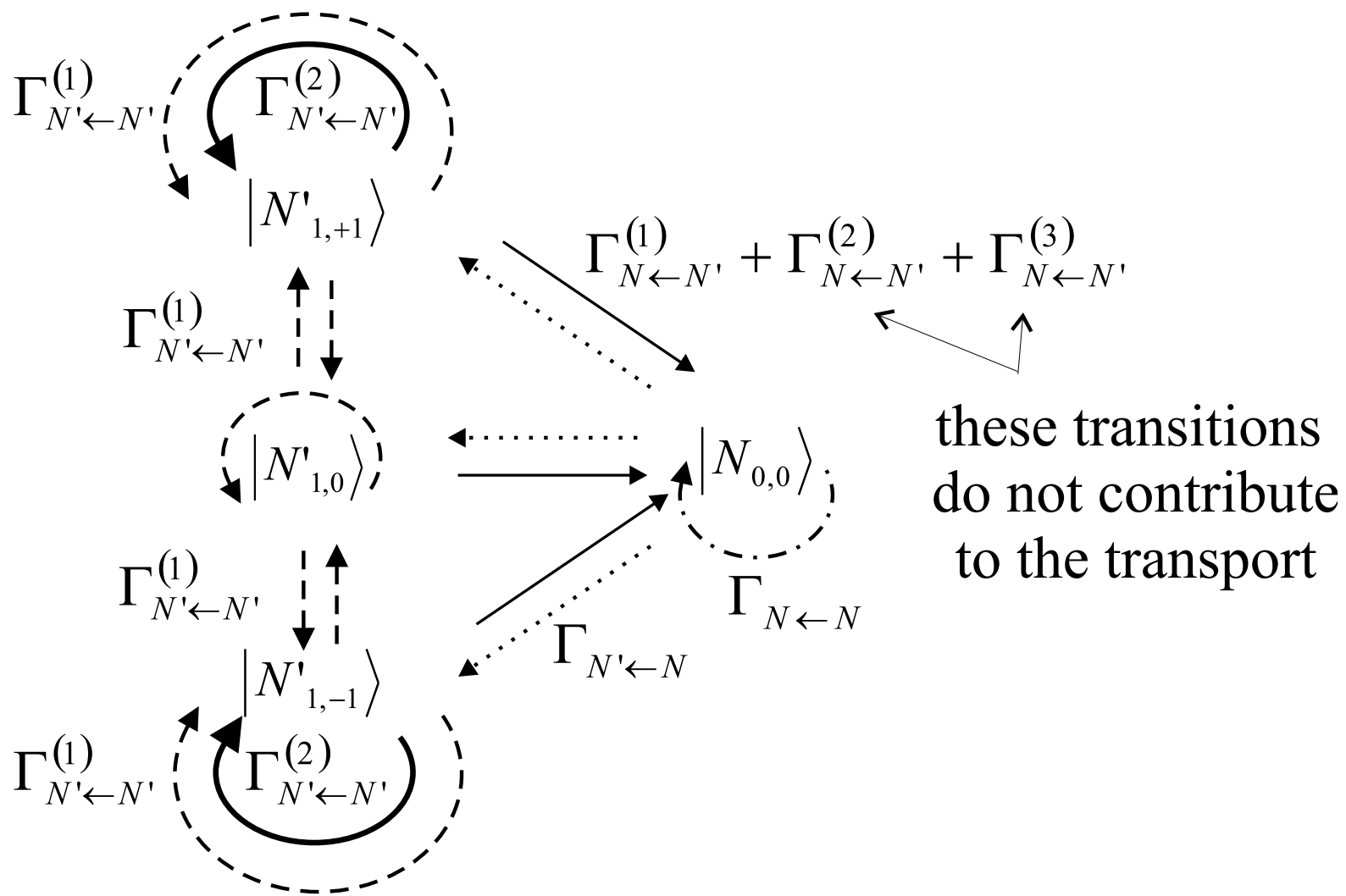
\end{document}